\documentclass[useAMS,usenatbib,usegraphicx,letterpaper]{mn2e}
\usepackage{times,amssymb,hyperref,aas_macros}

\usepackage[textwidth=480pt,textheight=680pt,layoutvoffset=0.5cm]{geometry}

\begin{document}

\title[Kuiper Belts in M-type Planet-host Systems]{Kuiper Belt Analogues
  in Nearby M-type Planet-host Systems}

\author[G. M. Kennedy et al.]{G. M. Kennedy\thanks{Email:
    \href{mailto:g.kennedy@warwick.ac.uk}{g.kennedy@warwick.ac.uk}}$^1$,
  G. Bryden$^2$,
  D. Ardila$^{3,4}$,
  C. Eiroa$^5$,
  J.-F. Lestrade$^6$,
  J. P. Marshall$^7$,
\newauthor
  B. C. Matthews$^{8,9}$,
  A. Moro-Martin$^{10}$, 
  M. C. Wyatt$^{11}$ \\
  $^1$ Department of Physics, University of Warwick, Gibbet Hill Road,
  Coventry, CV4 7AL, UK \\
  $^2$ Jet Propulsion Laboratory, California Institute of Technology,
  4800 Oak Grove Drive, Pasadena, CA 91109, USA \\
  $^3$ NASA Herschel Science Center, California Institute of Technology,
  MC 100-22, Pasadena, CA 91125, USA \\
  $^4$ The Aerospace Corporation, M2-266, El Segundo, CA 90245, USA \\
  $^5$ Dpto. Fisica Teorica, Universidad Autonoma de Madrid,
  Cantoblanco, 28049 Madrid, Spain \\
  $^6$ Observatoire de Paris - LERMA, CNRS, 61 Av. de l'Observatoire,
  75014, Paris, France \\
  $^7$ Academia Sinica, Institute of Astronomy and Astrophysics, Taipei
  10617, Taiwan \\
  $^8$ National Research Council of Canada Herzberg Astronomy \&
  Astrophysics Programs, 5071 West Saanich Road, Victoria, BC, V9E 2E7,
  Canada \\
  $^9$ Department of Physics \& Astronomy, University of Victoria, 3800
  Finnerty Road, Victoria, BC, V8P 5C2, Canada \\
  $^{10}$ Space Telescope Science Institute, 3700 San Martin Dr,
  Baltimore, MD 21218, USA \\
  $^{11}$ Institute of Astronomy, University of Cambridge, Madingley
  Road, Cambridge CB3 0HA, UK }
\maketitle

\begin{abstract}
  We present the results of a \emph{Herschel} survey of 21 late-type
  stars that host planets discovered by the radial velocity
  technique. The aims were to discover new disks in these systems and to
  search for any correlation between planet presence and disk
  properties. In addition to the known disk around GJ~581, we report the
  discovery of two new disks, in the GJ~433 and GJ~649 systems. Our
  sample therefore yields a disk detection rate of 14\%, higher than the
  detection rate of 1.2\% among our control sample of DEBRIS M-type
  stars with 98\% confidence. Further analysis however shows that the
  disk sensitivity in the control sample is about a factor of two lower
  in fractional luminosity than for our survey, lowering the
  significance of any correlation between planet presence and disk
  brightness below 98\%. In terms of their specific architectures, the
  disk around GJ~433 lies at a radius somewhere between 1 and 30au. The
  disk around GJ~649 lies somewhere between 6 and 30au, but is
  marginally resolved and appears more consistent with an edge-on
  inclination. In both cases the disks probably lie well beyond where
  the known planets reside (0.06-1.1au), but the lack of radial velocity
  sensitivity at larger separations allows for unseen Saturn-mass
  planets to orbit out to $\sim$5au, and more massive planets beyond
  5au. The layout of these M-type systems appears similar to Sun-like
  star + disk systems with low-mass planets.
\end{abstract}


\section{Introduction}\label{s:intro}

It is now well established that planet formation processes are robust,
and proceed around stars of a wide range of masses. At the higher mass
end, planets have been discovered around evolved stars with masses up to
three times the Sun's
\citep[e.g.][]{2005A&A...437L..31S,2007ApJ...665..785J,2015A&A...574A.116R}. At
the lower mass end the results have been equally impressive, with
planets discovered around objects ten times less massive than the Sun,
and whose luminosity is a thousand times weaker
\citep[e.g.][]{2016Natur.533..221G,2016Natur.536..437A}. This wide mass
range provides a unique way to study planet formation processes, and has
shown that while the occurence rate of giant planets increases towards
higher mass stars
\citep{2007ApJ...670..833J,2010PASP..122..905J,2015A&A...574A.116R}, the
converse is true for the frequency of Earth to Neptune-mass planets
\citep{2015ApJ...798..112M}.

In tandem with these searches, observations that seek to detect the
building blocks of these planets have also been conducted. These mid and
far-infrared (IR) surveys detect `debris disks', the collections of
small dust particles that are seen to orbit other stars (the `dust'
comprises various constituents, such as silicates, ice, and organic
compounds). Since their discovery in the 1980's, a growing body of
evidence has shown that they can be interpreted as circumstellar disks
made up of bodies ranging from
$\sim$$\mu$m to many km in size; while the observations only detect
$\mu$m to mm-size particles, the lifetime of these particles is commonly
shorter than the age of the host star, leading to the conclusion that
they must be replenished through the collisional destruction of a mass
reservoir of larger planetesimals
\citep[e.g.][]{1993prpl.conf.1253B}. For main-sequence stars this
paradigm is generally accepted, so in terms of the dust having an origin
in collisions between larger bodies, debris disks can be genuinely
thought of as analogues of the Solar System's Asteroid and Kuiper
belts. A key unknown is how the planetesimals acquire high enough
relative velocities for their collisions to be destructive; while it is
possible that planets excite these velocities
\citep{2009MNRAS.399.1403M}, it may be a natural outcome upon emergence
from the gas rich phase of evolution, or the planetesimals may `stir'
themselves \citep[e.g.][]{2004AJ....127..513K}, in which case planets
are not necessarily needed in order for debris disks to exist.

However, it is well known that the Solar System planets play an
important role in sculpting the Asteroid and Kuiper belts, two examples
being the presence of the Kirkwood gaps and the capture of Pluto into
2:3 mean motion resonance by Neptune \citep{1993Natur.365..819M}. In
attempts to make analogous link in other planetary systems, hypotheses
that connect the properties of the disks and planets have been
developed, and vary in complexity. The most basic is that some systems
are simply `better' at forming large bodies (whether those bodies be
planetesimals or planets), and more detailed models suggest that the
outcomes depend on whether planetary instabilities occurred
\citep{2011A&A...530A..62R}. As with planets, merely detecting these
belts is challenging, so quantifying the connection between the planets
and disks in these systems is typically limited to searching for
correlations between their basic properties \citep[such as disk
brightness,
e.g.][]{2009ApJ...700L..73K,2009ApJ...705.1226B,2012MNRAS.424.1206W,2014A&A...565A..15M,2015AJ....149...86W,2015ApJ...801..143M}. Ultimately,
these searches yielded a significant correlation between the presence of
radial velocity planets and the brightness of debris disks around
Sun-like stars \citep{2014prpl.conf..521M}. This trend is unfortunately
not strong, so while splitting the sample to look for trends among
sub-samples (e.g. as a function of planet mass) yields tentative trends
\citep[e.g.][]{2012MNRAS.424.1206W} it also lowers the significance.
Thus, while there is evidence that some Sun-like stars are indeed better
at forming disks and planets than other, the origin of this correlation
remains unclear.

In the case of low-mass stars the challenge of finding connections
between the planet and disk populations is even greater; for disks at
the typical radial distances of a few tens of astronomical units, the
low stellar luminosities do not heat the dust to temperatures greater
than about 50K. While the Stefan-Boltzmann law therefore limits the
luminosity of these disks, the low temperatures further hinder detection
because discoveries must be made at far infrared and millimeter
wavelengths
\citep[e.g.][]{2006A&A...460..733L,2012A&A...548A..86L}. Thus, it is not
particularly surprising that efforts to discover debris disks around
late-type stars at mid-infrared wavelengths have often been unsuccessful
\citep[e.g.][]{2007ApJ...667..527G,2012A&A...548A.105A}. Further, the
sensitivity of surveys is normally such that the non-detections are not
sufficiently constraining to rule out disks that have similar properties
to those that are known to orbit Sun-like stars
\citep{2007ApJ...667..527G,2014A&A...565A..58M}.

In this paper, we present far infrared \emph{Herschel}\footnote{Herschel
  is an ESA space observatory with science instruments provided by
  European-led Principal Investigator consortia and with important
  participation from NASA} \citep{2010A&A...518L...1P} observations that
aim to detect Kuiper belt analogues around a sample of 21 nearby late K
and M-type stars that host planets discovered by the radial velocity
technique. The primary aim is to search for a correlation between the
presence of planets and the brightness of disks, and secondary aims are
to detect new disks that may be amenable to further detailed
investigation, and to obtain more sensitive observations than were
possible with larger surveys. We present the sample and observations in
section \ref{s:obs}, discuss the results in section \ref{s:disc}, and
summarise and conclude in section \ref{s:conc}.

\section{Sample and Observations}\label{s:obs}

\begin{table}
  \begin{center}
    \caption{PACS observations of 16 targets taken as part of our
      programme (OT2\_gbryden\_2). OD is the Herschel Observing Day, and
      Reps is the number of repeats of a standard PACS mini scan-map
      used to reach the desired sensitivity.}\label{tab:obs}
  \begin{tabular}{llll}
    \hline
    Name & ObsIDs & OD & Reps \\
    \hline
    GJ 176 & 1342250278/279 & 1202 & 6\\
    GJ 179 & 1342250276/277 & 1202 & 6\\
    GJ 317 & 1342253029/030 & 1245 & 6\\
    GJ 3634 & 1342257175/176 & 1310 & 6\\
    GJ 370 & 1342256997/998 & 1308 & 6\\
    GJ 433 & 1342257567/568 & 1316 & 6\\
    GJ 1148 & 1342247393/394 & 1138 & 6\\
    GJ 436 & 1342247389/390 & 1138 & 6\\
    GJ 9425 & 1342249877/878 & 1194 & 6\\
    GJ 9482 & 1342248728/729 & 1170 & 6\\
    HIP 79431 & 1342262219/220 & 1355 & 6\\
    GJ 649 & 1342252819/820 & 1244 & 6\\
    GJ 1214 & 1342252011/012 & 1237 & 6\\
    GJ 674 & 1342252841/842 & 1244 & 6\\
    GJ 676 A & 1342243794/795 & 1058 & 6\\
    GJ 849 & 1342246764/765 & 1121 & 6\\
  \end{tabular}  
\end{center}
\end{table}

Our sample comprises nearly all low-mass planet-host stars within
20pc. Most stars are M spectral type, but we include three that are late
K types (GJ~370, GJ~9425 and GJ~9482). Not all systems in the final
sample were known to host planets at the time the observations were
proposed (2011 September), but some in which planets were subsequently
discovered were observed by the volume-limited DEBRIS Key Programme
\citep{2010A&A...518L.135M}. The final sample has 21 stars, 16 of which
were observed by \emph{Herschel} in this programme, and which are listed
in Table \ref{tab:obs}. Five more targets, GJ~15~A, GJ~581, GJ~687,
GJ~842, and GJ~876, were observed by the DEBRIS survey so are also
included in our sample \citep[see][for results for
GJ~581]{2012A&A...548A..86L}.

The sample does not include the planet host Proxima Centauri
\citep{2016Natur.536..437A}, as it was not observed by
\emph{Herschel}. While it has been suggested to host excess emission
arising from a debris disk \citep{2017arXiv171100578A}, these
observations use the Atacama Large Millimeter Array (ALMA) and this
system is therefore not easily integrated into our sample. Two of our
targets are possible wide binaries; GJ~15~A (NLTT~919) is a common
proper motion pair at a projected separation of 35\arcsec\ with NLTT~923
\citep{2004ApJS..150..455G}, and GJ~676~A has a wide common proper
motion companion (GJ~676~B) at a projected separation of 50\arcsec\
\citep{1994RMxAA..28...43P}. We do not expect the planetary systems to
be affected seriously by these companions, so retain them in our sample.

\begin{figure*}
  \begin{center}
    \hspace{-0.4cm} \includegraphics[width=0.275\textwidth]{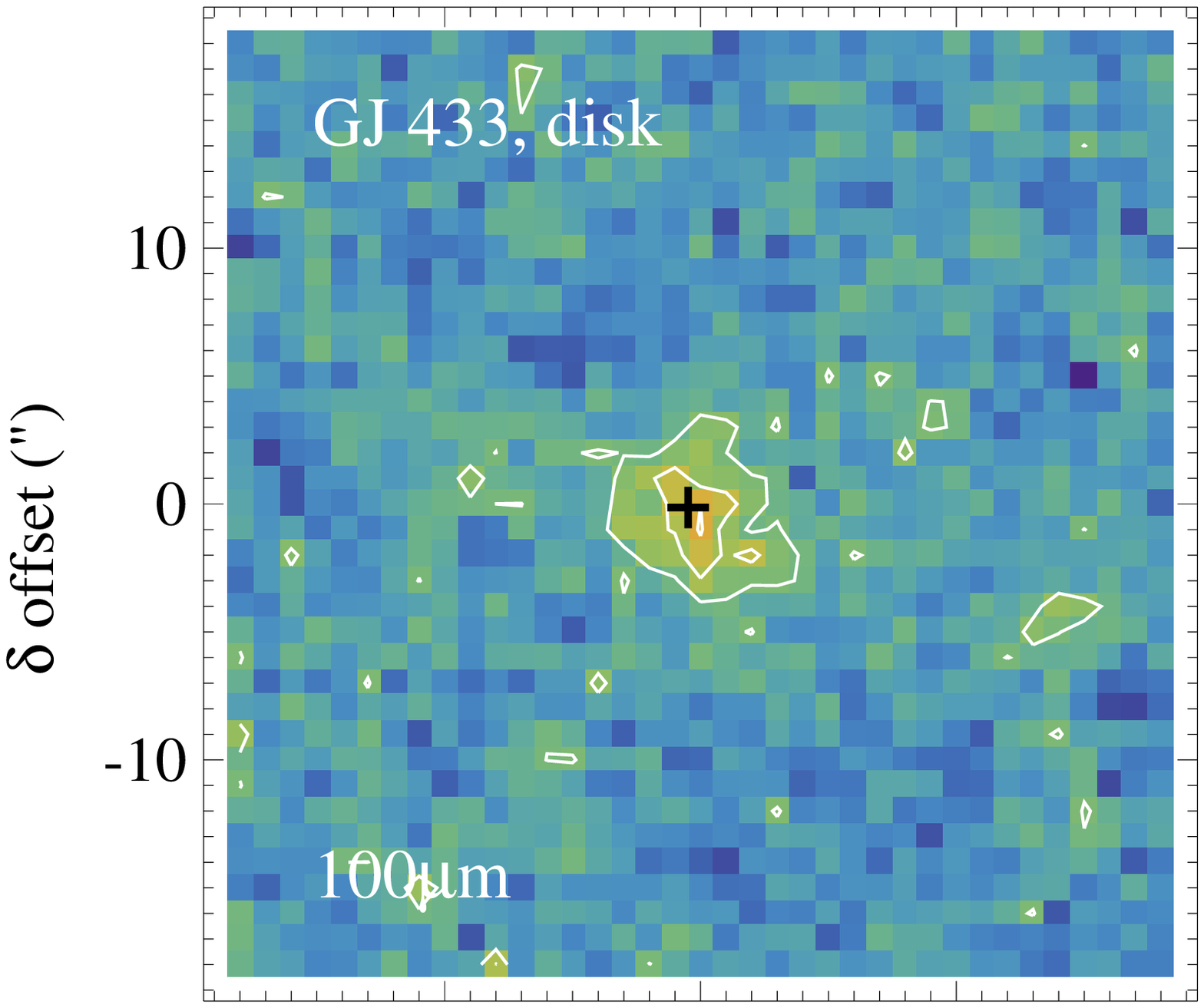}
    \hspace{-1.8cm} \includegraphics[width=0.275\textwidth]{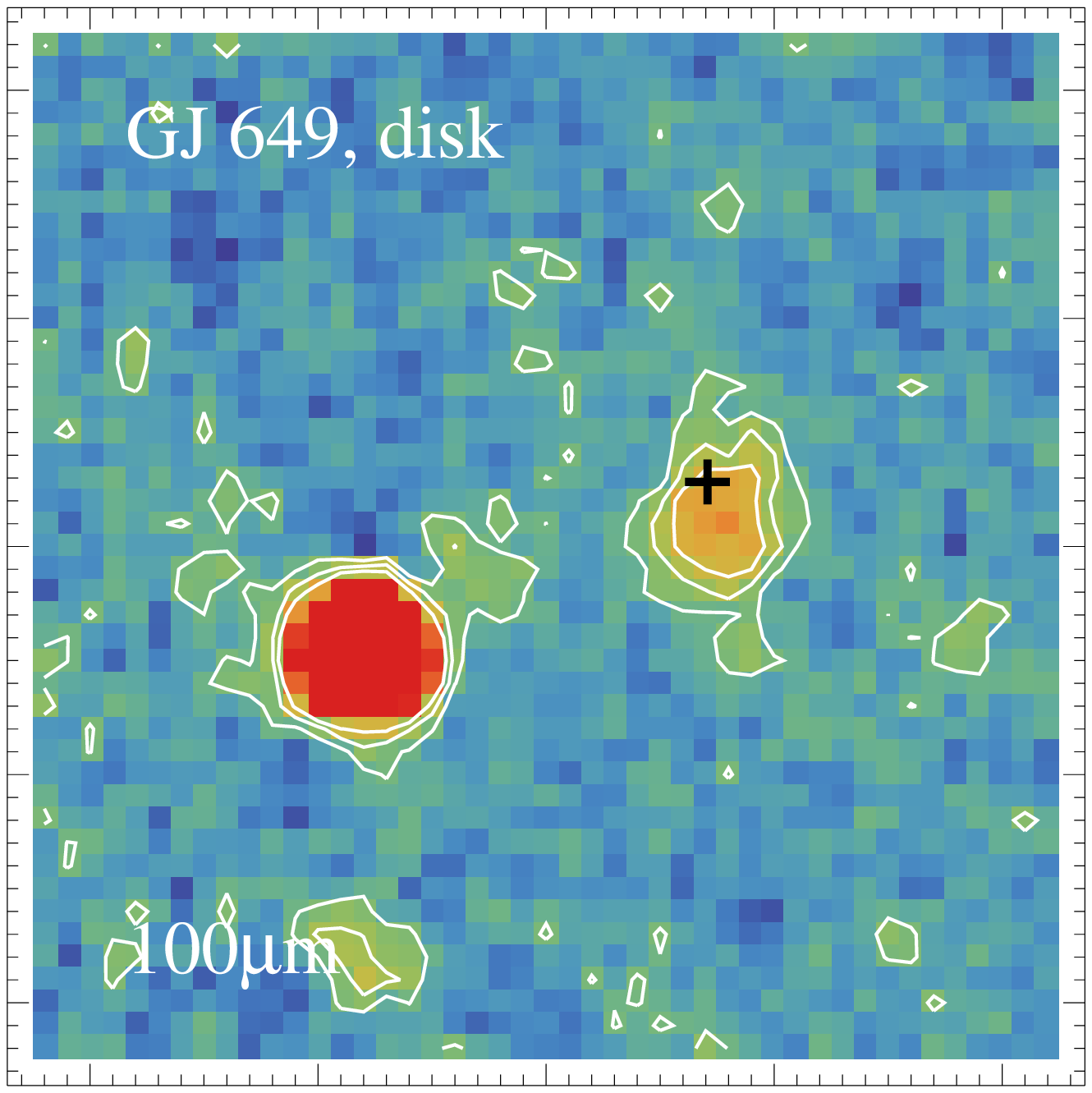}
    \hspace{-1.8cm} \includegraphics[width=0.275\textwidth]{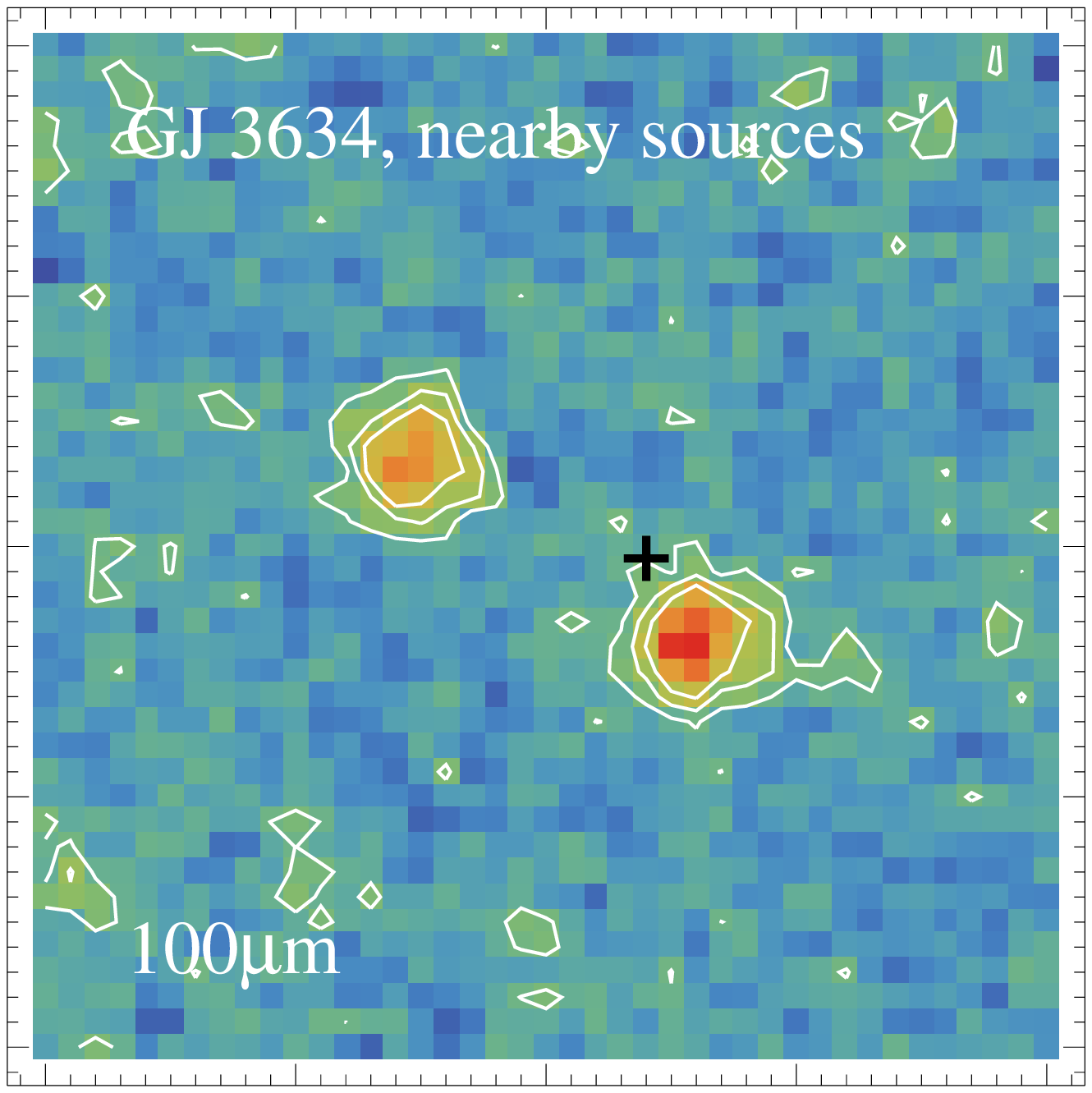}
    \hspace{-1.8cm} \includegraphics[width=0.275\textwidth]{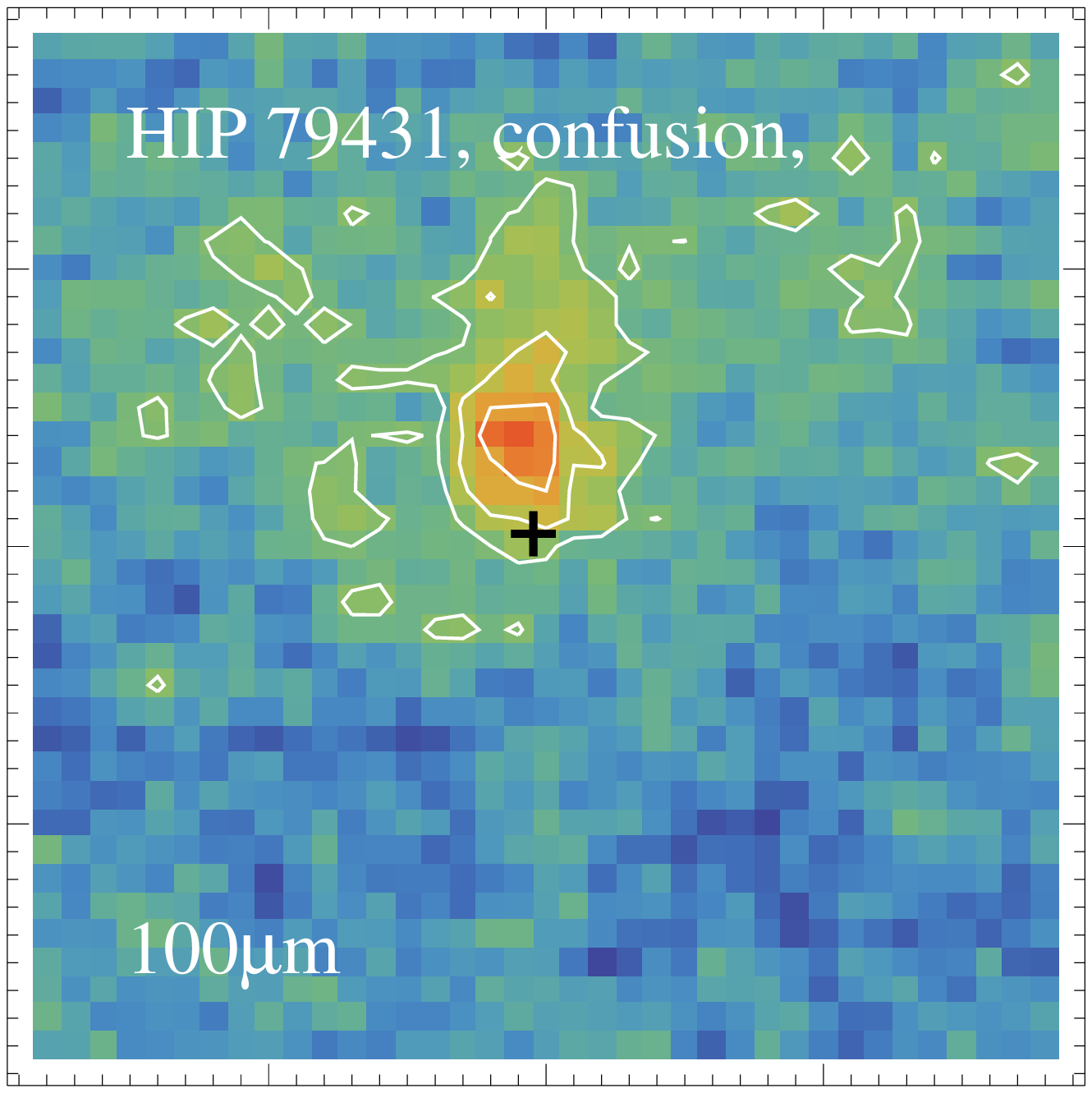}
    \hspace{-1.8cm} \includegraphics[width=0.275\textwidth]{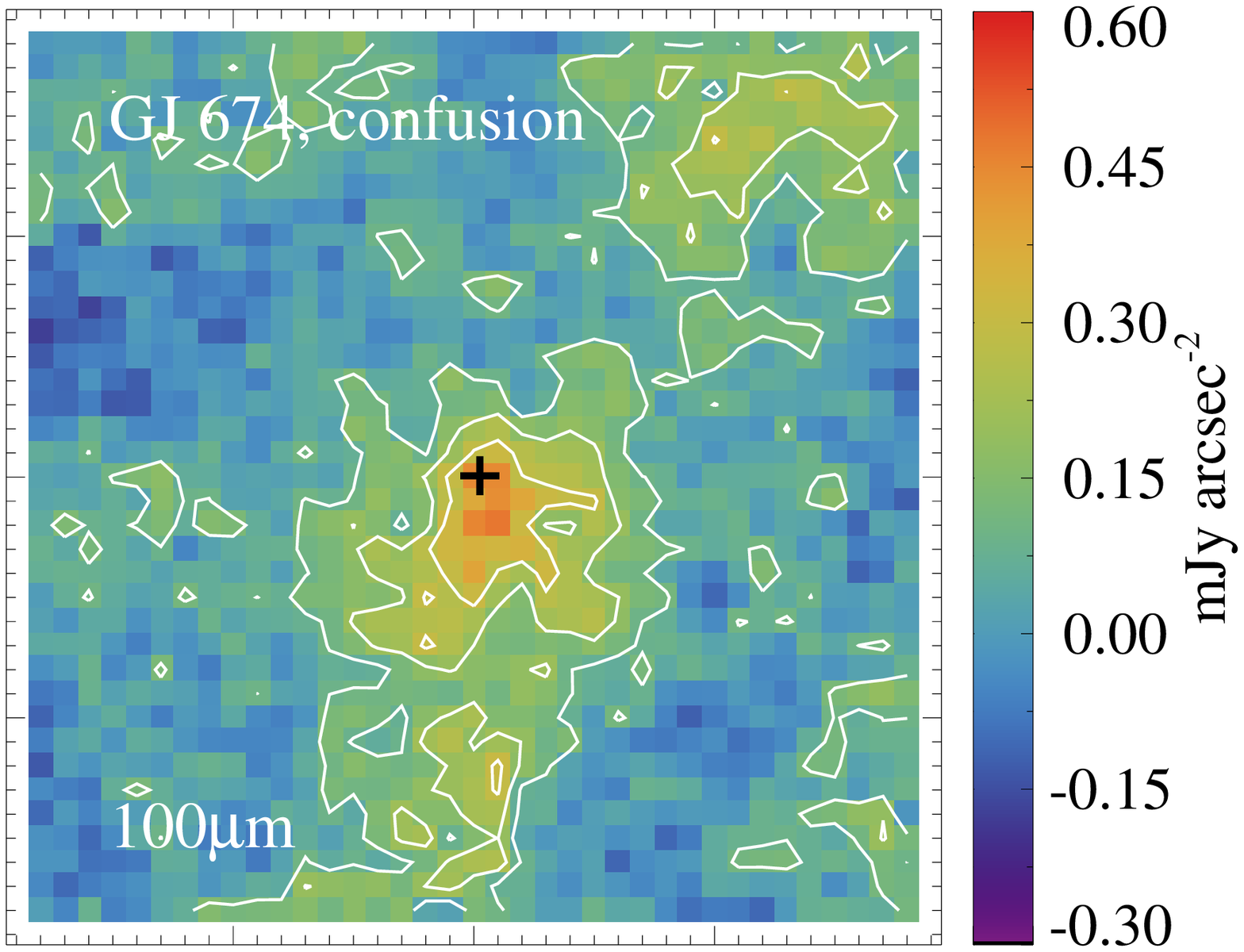} \\
    \vspace{-0.95cm}
    \hspace{-0.4cm} \includegraphics[width=0.275\textwidth]{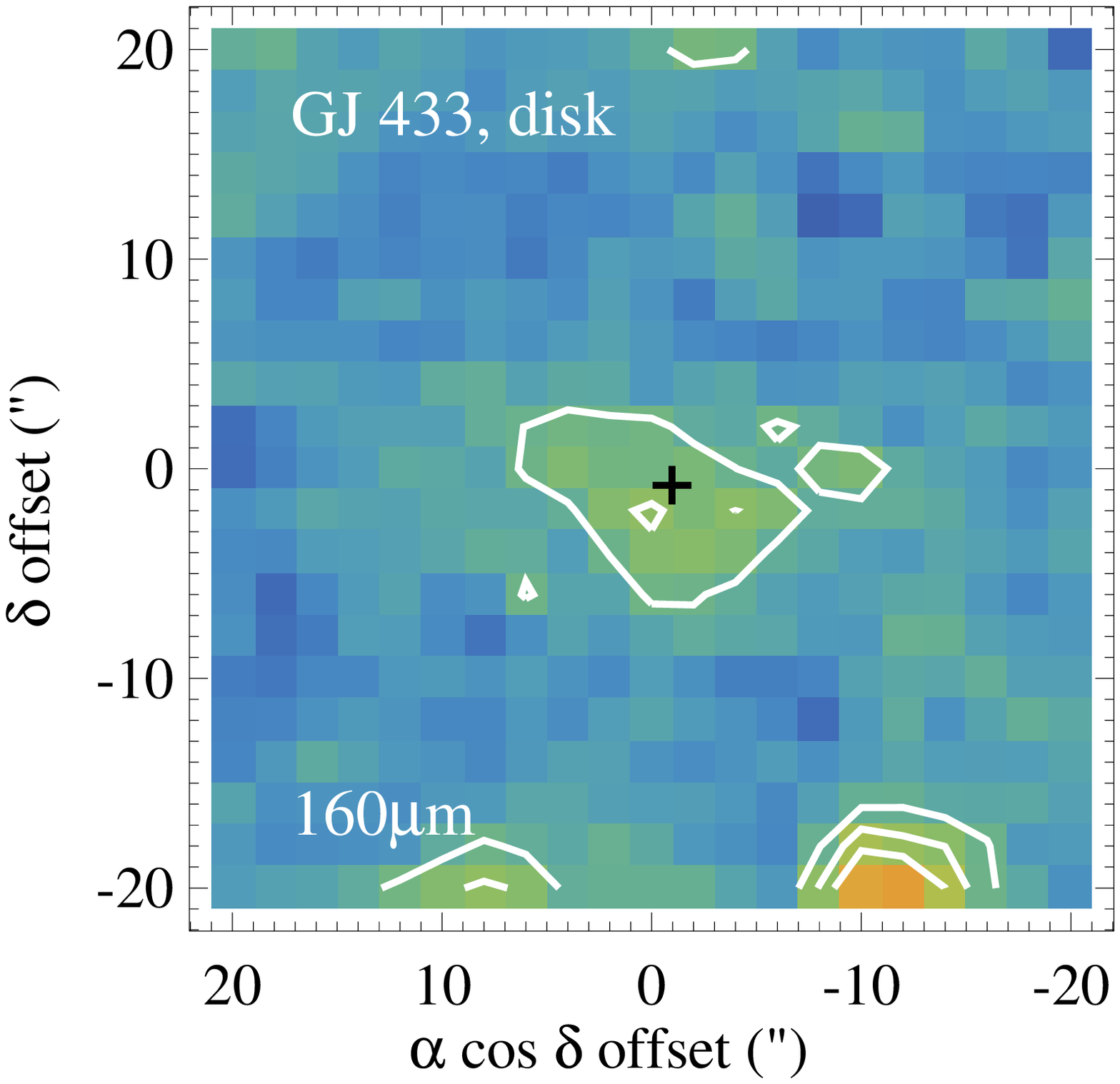}
    \hspace{-1.8cm} \includegraphics[width=0.275\textwidth]{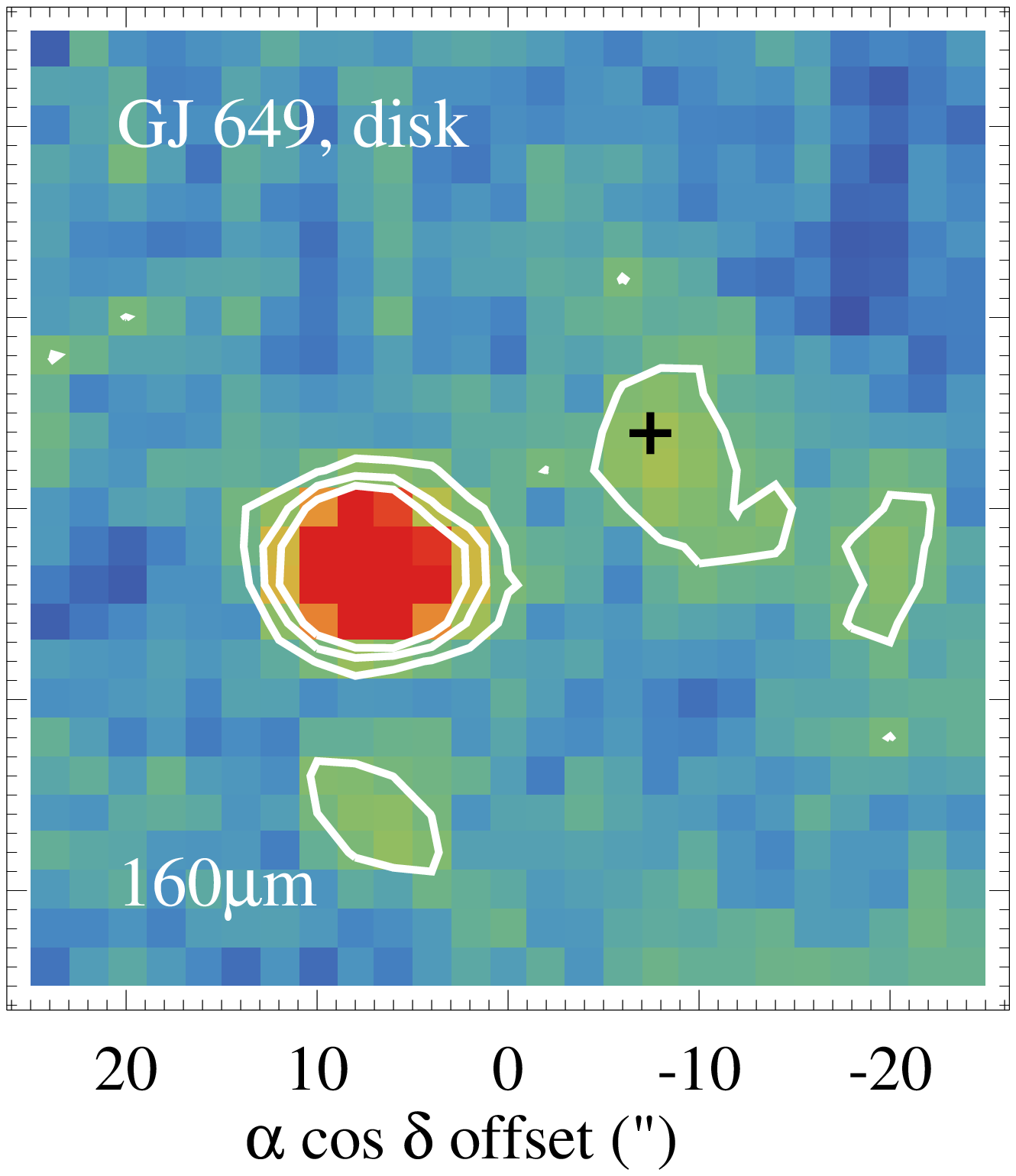}
    \hspace{-1.8cm} \includegraphics[width=0.275\textwidth]{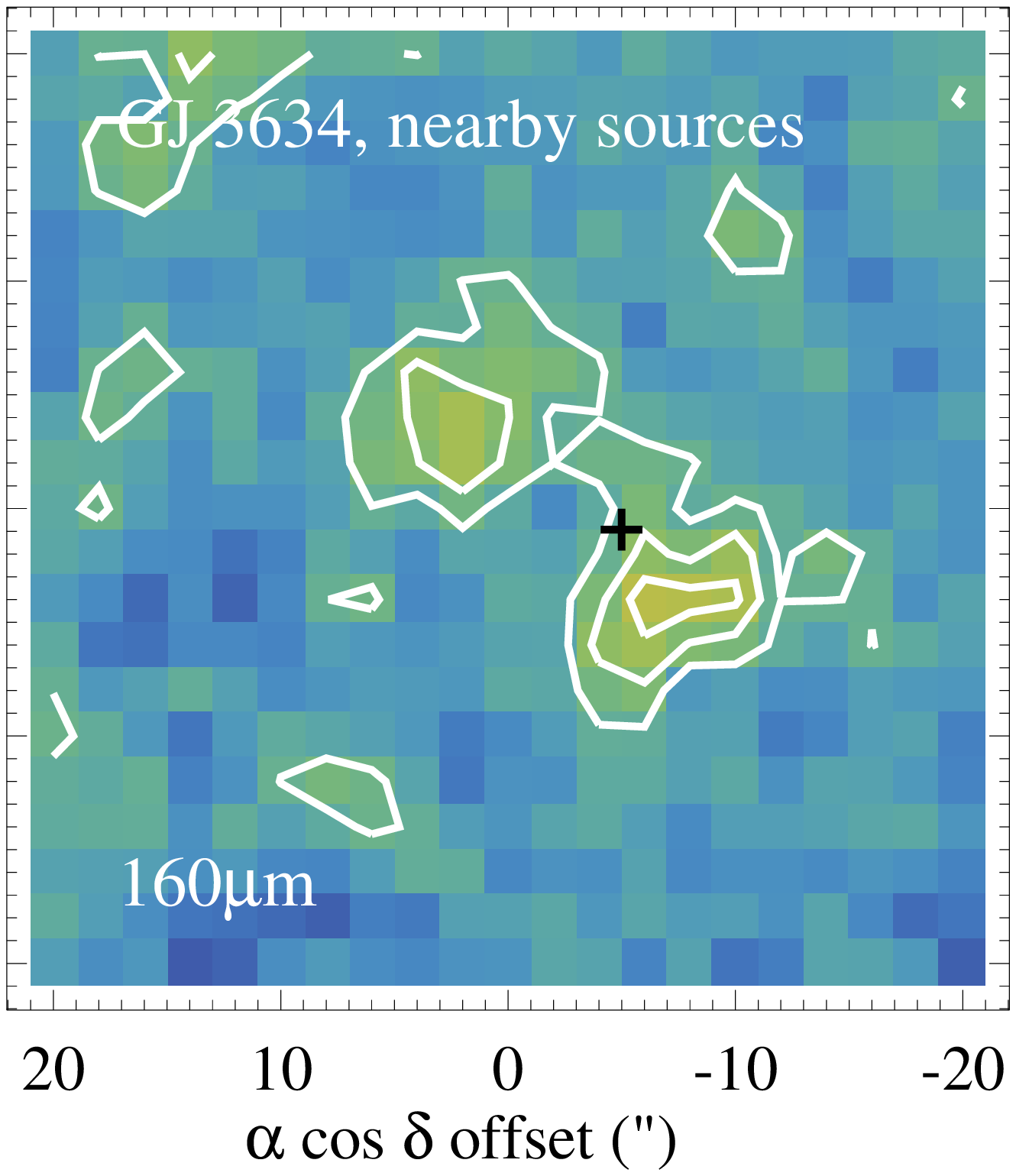}
    \hspace{-1.8cm} \includegraphics[width=0.275\textwidth]{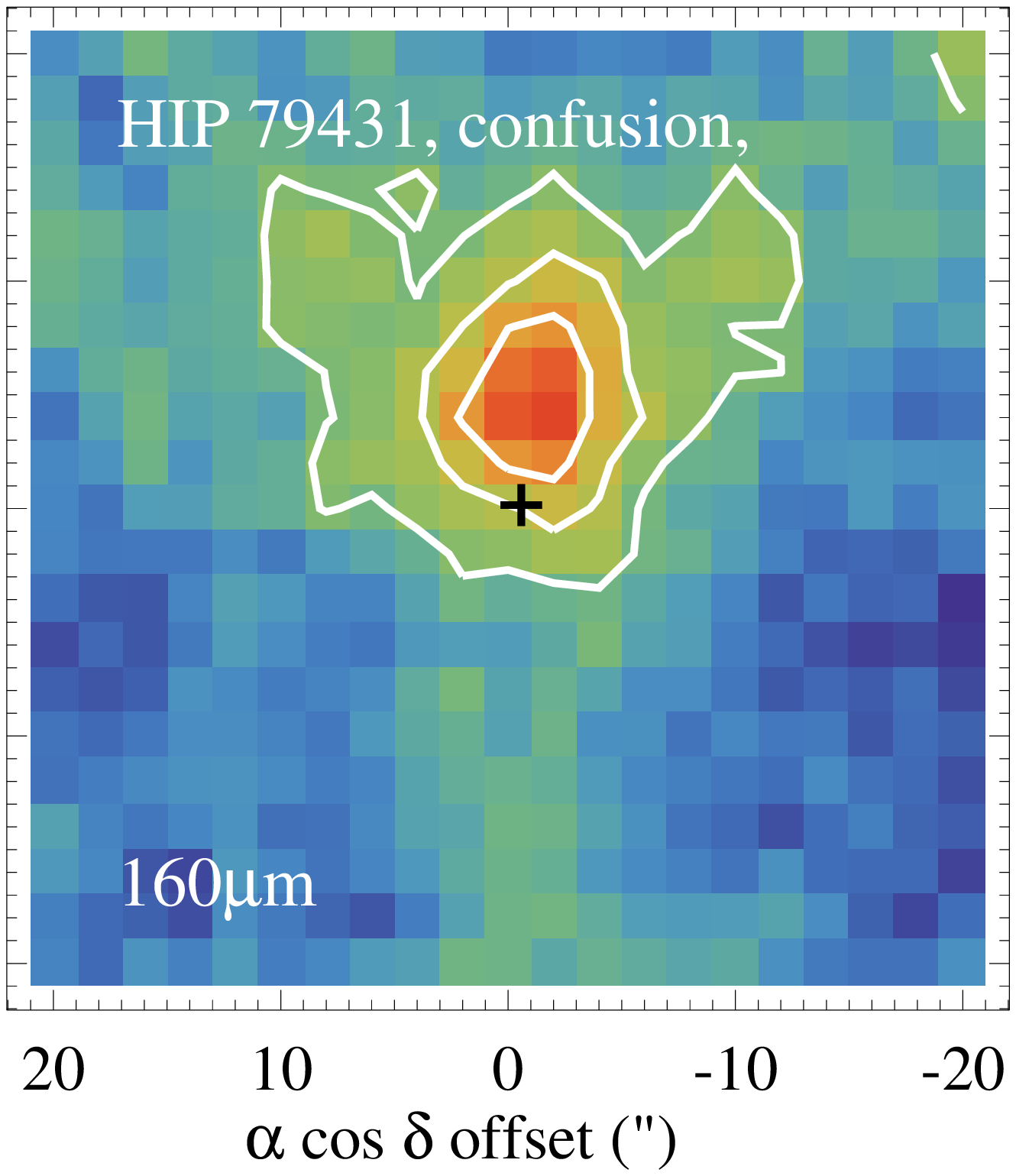}
    \hspace{-1.8cm} \includegraphics[width=0.275\textwidth]{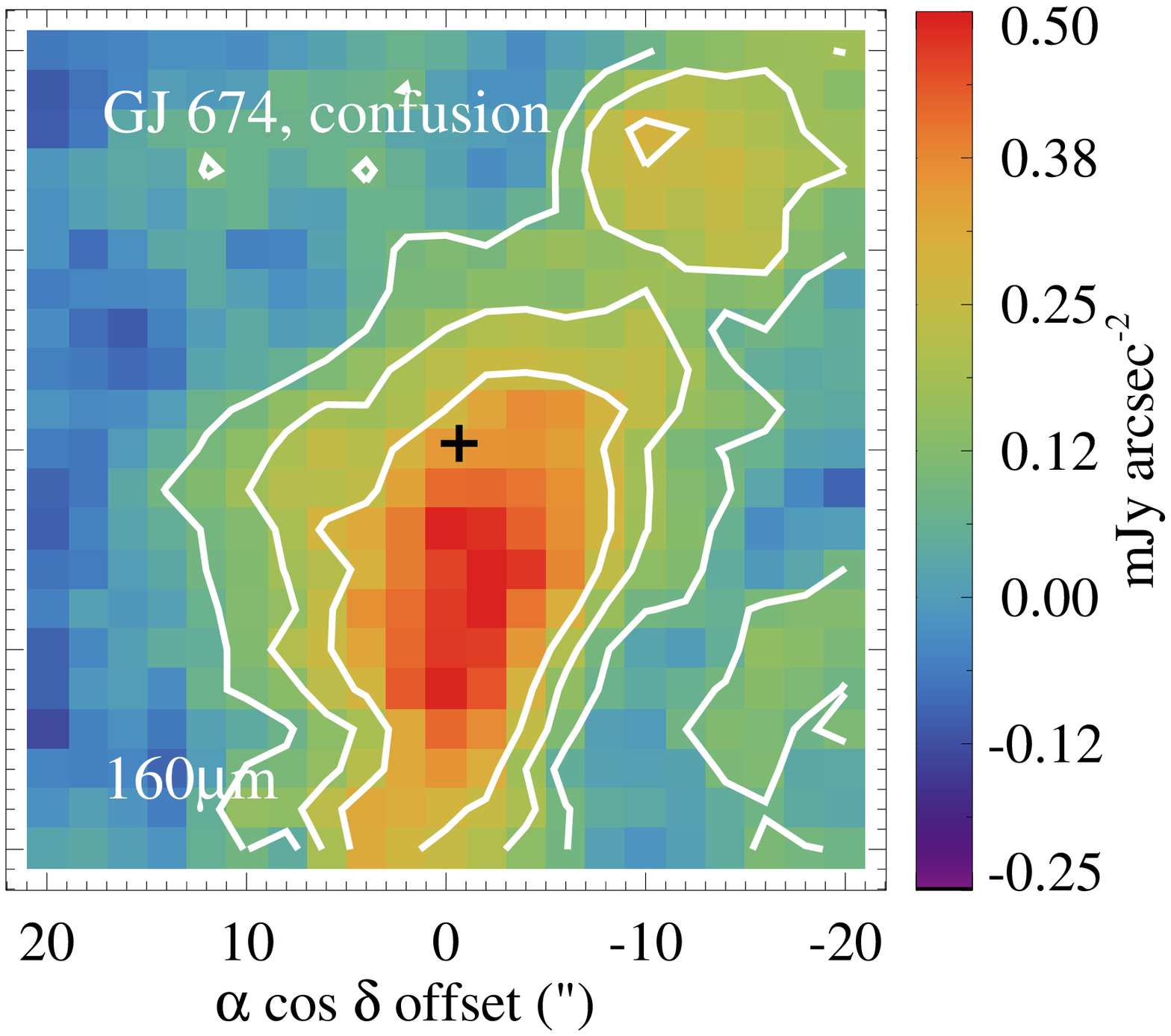}
    \caption{\emph{Herschel} images of the two targets found here to
      host debris disks (GJ~433 and GJ~649, in the left two columns),
      and the three targets for which excess emission near the star was
      seen, but which was assumed not to be associated with the star in
      question (right three columns). In each panel the black cross
      marks the estimated stellar position at the time of
      observation. Each image is centered either on the star, or in the
      case of GJ~649 and GJ~3634 between the two visible source
      detections. White contours are at 2, 4, and 6 times the 1$\sigma$
      noise level in each image. The disk around GJ~649 appears to be
      marginally resolved; see Figure \ref{fig:im2}.}\label{fig:im}
  \end{center}
\end{figure*}

The targets were observed using the Photodetector Array Camera and
Spectrometer \citep[PACS,][]{2010A&A...518L...2P}, using the so-called
`mini-scan map' mode. A series of ten parallel scans with a separation
of 4\arcsec\ are taken to make a single map, which is repeated six times
to build up the signal. One such sequence coresponds to a single
observation ID number, or ObsID. The observatory is then rotated by
40$^\circ$, and the sequence repeated, to provide some robustness to
striping artefacts and low-frequency noise. The total integration time
for each source is 56 minutes. For our observations the noise level at
100$\mu$m was typically 1mJy, while observations carried out by DEBRIS
(integration time of 15 minutes) had fewer repeats and a noise level
nearer 2mJy. The images used in the analysis are the standard `level
2.5' observatory products obtained from the Herschel Science
Archive,\footnote{\href{http://archives.esac.esa.int/hsa/whsa/}{http://archives.esac.esa.int/hsa/whsa/}}
which combine the two observing sequences (ObsIDs) into a single image.

Photometry $F_{\rm obs}$ for each source was extracted using point
spread function (PSF) fitting. Observations of the calibration star
$\gamma$ Dra, again level 2.5 observatory products, were used as PSFs,
which were rotated to a position angle appropriate for each
observation. The fitting was done at 100 and 160$\mu$m simultaneously,
so the four free parameters in each fit were a position common to both
wavelengths, and two fluxes (i.e. $F_{100}$ and
$F_{160}$). Uncertainties $\sigma_{100}$ and $\sigma_{160}$ were
estimated by measuring the flux in apertures at hundreds of random
locations near the center of the images; this method was found to be
more reliable and provide more realistic flux distributions than
attempting to fit PSFs at random locations. The apertures were chosen to
be those optimal for source extraction (5 and 8\arcsec~for 100 and
160$\mu$m respectively, derived using calibration observations). In the
case of GJ~649 there is evidence that the source (i.e. disk) is
marginally resolved (see Figure \ref{fig:im}), so the flux for this
source at 100$\mu$m is measured using an aperture radius of 10\arcsec,
and the uncertainty estimated as above but with 10\arcsec\
apertures. The results of the source extraction are summarised in Table
\ref{tab:fluxes}, and the results for a few problematic sources are
described in more detail below.

To assess whether each star shows the infrared excess that is indicative
of a debris disk requires an estimate of the flux density expected from
the stellar photosphere $F_\star$ at the PACS wavelengths. These
estimates are made by fitting stellar photosphere models to optical and
near-IR photometry. The method has been described elsewhere, and for
example has been used for the DEBRIS survey and shown to provide
photospheric fluxes that are sufficiently precise that the detection of
excesses is limited by the \emph{Herschel} photometry, not the
photosphere models \citep[i.e.
$\sigma_{\rm obs} >
\sigma_\star$,][]{2012MNRAS.426.2115K,2012MNRAS.421.2264K}. While
photospheric models for late-type stars are less precise than for
earlier types (e.g. because of uncertain molecular opacity), the flux of
many of our target stars is predicted to be near our noise level and the
models are not a limiting factor. The photospheric predictions at 100
and 160$\mu$m are given in Table~\ref{tab:fluxes}.

The significance of any excess is then given in each PACS bandpass by
$\chi = (F_{\rm obs} - F_\star) / \sqrt{\sigma_{\rm obs}^2 +
  \sigma_\star^2}$, where $\chi>3$ is taken to be a significant
excess. To summarise the observational results; in addition to the disk
known to orbit GJ~581, we find two new systems that show strong evidence
for infrared excesses: GJ~433 and GJ~649, whose images are shown in
Figure \ref{fig:im}.


\begin{table*}
  \caption{The 21 stars in our sample, comprising 16 stars observed in programme
    OT2\_gbryden\_2, and five stars observed in  programme
    KPOT\_bmatthew\_1 (DEBRIS): GJ~15, GJ~581 \citep[multiple observations,
    see][]{2012A&A...548A..86L}, GJ~687, GJ~832, and GJ~876. We have not
    reported flux densities for the two strongly confused sources,
    HIP~79431 and GJ~674.}\label{tab:fluxes} 
  \begin{tabular}{lrlrrrrrrrrrl}
    \hline
    GJ & HIP no. & SpTy & Dist &
    $F_{\star,100}$ & $F_{\rm 100}$ & $\sigma_{\rm 100}$ & $\chi_{\rm 100}$ &
    $F_{\star,160}$ & $F_{\rm 160}$ & $\sigma_{\rm 160}$ & $\chi_{\rm 160}$ &
    Notes \\
    & & & (pc) &
    (mJy) & (mJy) & (mJy) & &
    (mJy) & (mJy) & (mJy) & & \\
    \hline
GJ 15 A & 1475 & M2V & 3.6 & 15.3 & 14.9 & 2.2 & -0.2 & 5.9 & 13.2 & 3.0 & 2.4 & Photosphere at 100$\mu$m \\
GJ 176 & 21932 & M2.5V & 9.4 & 4.1 & 3.6 & 1.6 & -0.3 & 1.6 & -3.6 & 6.4 & -0.8 & No detection \\
GJ 179 & 22627 & M2V & 12.4 & 1.3 & -1.6 & 1.2 & -2.5 & 0.5 & -4.7 & 2.8 & -1.9 & No detection \\
GJ 317 & - & M3.5V & 15.3 & 1.1 & 3.2 & 1.1 & 1.9 & 0.4 & 4.9 & 2.2 & 2.0 & No detection \\
GJ 370 & 48331 & K6Vk: & 11.3 & 5.7 & 6.9 & 0.9 & 1.3 & 2.2 & -5.2 & 3.2 & -2.3 & Photosphere at 100$\mu$m \\
GJ 3634 & - & M2.5 & 19.8 & 0.7 & -0.9 & 1.2 & -1.3 & 0.3 & 3.0 & 2.8 & 1.0 & Detection at 6\arcsec~SW \\
\textbf{GJ 433} & 56528 & M2V & 9.1 & 3.9 & 11.9 & 1.3 & 6.2 & 1.5 & 13.9 & 4.3 & 2.9 & Excess detection \\
GJ 1148 & 57050 & M4.0Ve & 11.1 & 1.5 & 1.4 & 1.0 & -0.1 & 0.6 & -2.1 & 3.3 & -0.8 & No detection \\
GJ 436 & 57087 & M3V & 9.7 & 2.4 & 3.4 & 1.1 & 0.9 & 0.9 & 4.3 & 2.3 & 1.5 & No detection \\
GJ 9425 & 63833 & K9Vk: & 15.9 & 3.1 & -0.7 & 2.1 & -1.8 & 1.2 & -15.1 & 7.6 & -2.1 & No detection \\
GJ 9482 & 70849 & K7Vk & 23.6 & 1.0 & 1.6 & 1.4 & 0.5 & 0.4 & -4.1 & 3.4 & -1.3 & No detection \\
GJ 581 & 74995 & M3V & 6.3 & 3.8 & 21.8 & 1.5 & 11.8 & 1.5 & 22.4 & 5.0 & 4.2 & Excess {\citet{2012A&A...548A..86L}} \\
- & 79431 & M3V & 14.4 & 1.7 & - & - & - & 0.6 & - & - & - & Extended detection at 5\arcsec~N \\
\textbf{GJ 649} & 83043 & M2V & 10.4 & 3.6 & 22.6 & 2.4 & 7.9 & 1.4 & 16.3 & 5.2 & 2.9 & Excess detection, extended? \\
GJ 1214 & - & M4.5V & 14.6 & 0.3 & 0.9 & 1.1 & 0.6 & 0.1 & -0.1 & 2.3 & -0.1 & No detection, source 10\arcsec~W \\
GJ 674 & 85523 & M3V & 4.5 & 8.1 & - & - & - & 3.1 & - & - & - & Extended, high background \\
GJ 676 A & 85647 & M0V & 15.9 & 2.6 & 0.9 & 1.1 & -1.6 & 1.0 & 0.5 & 2.2 & -0.2 & No detection, source 10\arcsec~SW \\
GJ 687 & 86162 & M3.0V & 4.5 & 10.1 & 6.1 & 1.6 & -2.5 & 3.9 & 0.2 & 3.4 & -1.1 & No detection \\
GJ 832 & 106440 & M2/3V & 5.0 & 10.4 & 12.5 & 1.6 & 1.3 & 4.0 & 1.2 & 3.5 & -0.8 & Photosphere at 100$\mu$m \\
GJ 849 & 109388 & M3.5V & 8.8 & 4.3 & 4.2 & 1.2 & -0.1 & 1.7 & 3.6 & 1.8 & 1.1 & No detection \\
GJ 876 & 113020 & M3.5V & 4.7 & 8.1 & 6.5 & 1.6 & -1.0 & 3.2 & 6.5 & 3.5 & 0.9 & Photosphere at 100$\mu$m \\
    \hline
  \end{tabular}  
\end{table*}

Several other targets were also found to have emission at or near the
source position, but in these cases we do not believe the emission to be
associated with the star in question. These are shown in Figure
\ref{fig:im}.

\begin{itemize}

\item GJ~3634: A bright ($\sim$14mJy) source is seen 6\arcsec~SW of the
  expected position of GJ~3634. This offset is larger than expected
  given the $\sim$2\arcsec~1$\sigma$ pointing accuracy of
  \emph{Herschel}\footnote{\href{http://herschel.esac.esa.int/Docs/Herschel/html/ch02s04.html}{http://herschel.esac.esa.int/Docs/Herschel/html/ch02s04.html}}
  and our small sample size. By comparing the positions of several other
  sources detected in the 100 $\mu$m PACS image with the (optical) DSS2
  plates\footnote{\href{https://archive.stsci.edu/dss/}{https://archive.stsci.edu/dss/}}
  we found that three were almost perfectly coincident. Thus, we
  conclude that the 6\arcsec~offset seen is real, and that the PACS
  detection near GJ~3634 is not associated with this star.

\item HIP~79431: Extended structure is seen to the North of the stellar
  position, but the peak is 5\arcsec~away. Only one low S/N source was
  seen to be common between the PACS and DSS2 images, with perfect
  coincidence. The background as seen in IRAS and WISE images is complex
  and variable. We conclude that the large offset and high background
  mean that the detected source is unlikely to be associated with
  HIP~79431.

\item GJ~674: The background level around GJ~674 is significantly above
  zero. At 100 $\mu$m the flux in the image peaks at the position of
  GJ~674, but if a point source with the photospheric flux of GJ~674 is
  subtracted the background becomes uniform. Thus, we conclude that the
  image shows emission from the star GJ~674 superimposed on a
  non-negligible background, and that there is no evidence for excess
  emission from the star itself.

\end{itemize}

\section{Discussion}\label{s:disc}

Our survey finds two new excess detections, around the stars GJ~433 and
GJ~649. We first consider these detections as part of our sample, and
then take a closer look at the architecture of these two systems in more
detail.

\subsection{Planet - disk correlation}\label{ss:corr}

One of our goals was to test for a correlation between the brightness of
debris disks around low-mass stars and the presence of planets. That is,
all stars may host debris disks, but we can only detect those above a
given dust level, so we cannot test for a correlation between the
`existence' of planets and disks. The same is true for planet detection
of course, so we are in fact testing for a correlation between disks
above a given brightness threshold and planets above a given semi-major
axis vs. mass threshold (acknowledging that the star-to-star sensitivity
also varies). These thresholds are discussed below.

A significant correlation has been seen among Sun-like stars that host
radial velocity planets \citep{2014prpl.conf..521M}, and tentative
evidence that this trend is stronger for stars that host low-mass
planets was found among a small sample of nearby stars
\citep{2012MNRAS.424.1206W,2014A&A...565A..15M}. No clear trends were
seen in the volume-limited DEBRIS FGK-type sample considered by
\citet{2015ApJ...801..143M}, illustrating the tentative nature of the
latter trend, and that samples that do not specifically target
planet-host stars suffer from small numbers of planet hosts that limit
the power to discover trends.

Here, our sample comprises 21 planet-hosting low-mass stars that were
observed in search of IR excesses by \emph{Herschel}, for which three
were found to host disks. Thus, our detection rate is 14\%, but clearly
suffers from a small number of detections. As a control sample, we
consider the volume-limited DEBRIS M-type sample, which comprises 89
nearby stars \citep{2010MNRAS.403.1089P}. Of these, two were discovered
to host debris disks; the planet host GJ~581 \citep{2012A&A...548A..86L}
and the third star in the very wide Fomalhaut triple system, Fomalhaut~C
\citep{2014MNRAS.438L..96K}. We remove GJ~581 and the four other
planet-host stars from this sample, leaving 84 stars with one disk
detection, a rate of 1.2\%.

A Fisher's exact test to determine whether these two populations could
arise from the same underlying distribution yields a $p$-value of 0.025,
thus showing reasonable evidence that the planet-host stars have a
tendency to have more detectable (i.e. brighter) debris disks. The
Fomalhaut system is known to be relatively young, at 440Myr
\citep{2012ApJ...754L..20M}; if we were to assume that all of the planet
host systems are older than this and exclude Fomalhaut~C from the
control sample the $p$-value decreases to 0.01. However, we cannot be
sure that the planet-host stars are all older than the Fomalhaut system,
since for example GJ~674 may also be a relatively young system
\citep{2007A&A...474..293B}.



\begin{figure*}
  \begin{center}
    \hspace{-0.5cm} \includegraphics[width=0.48\textwidth]{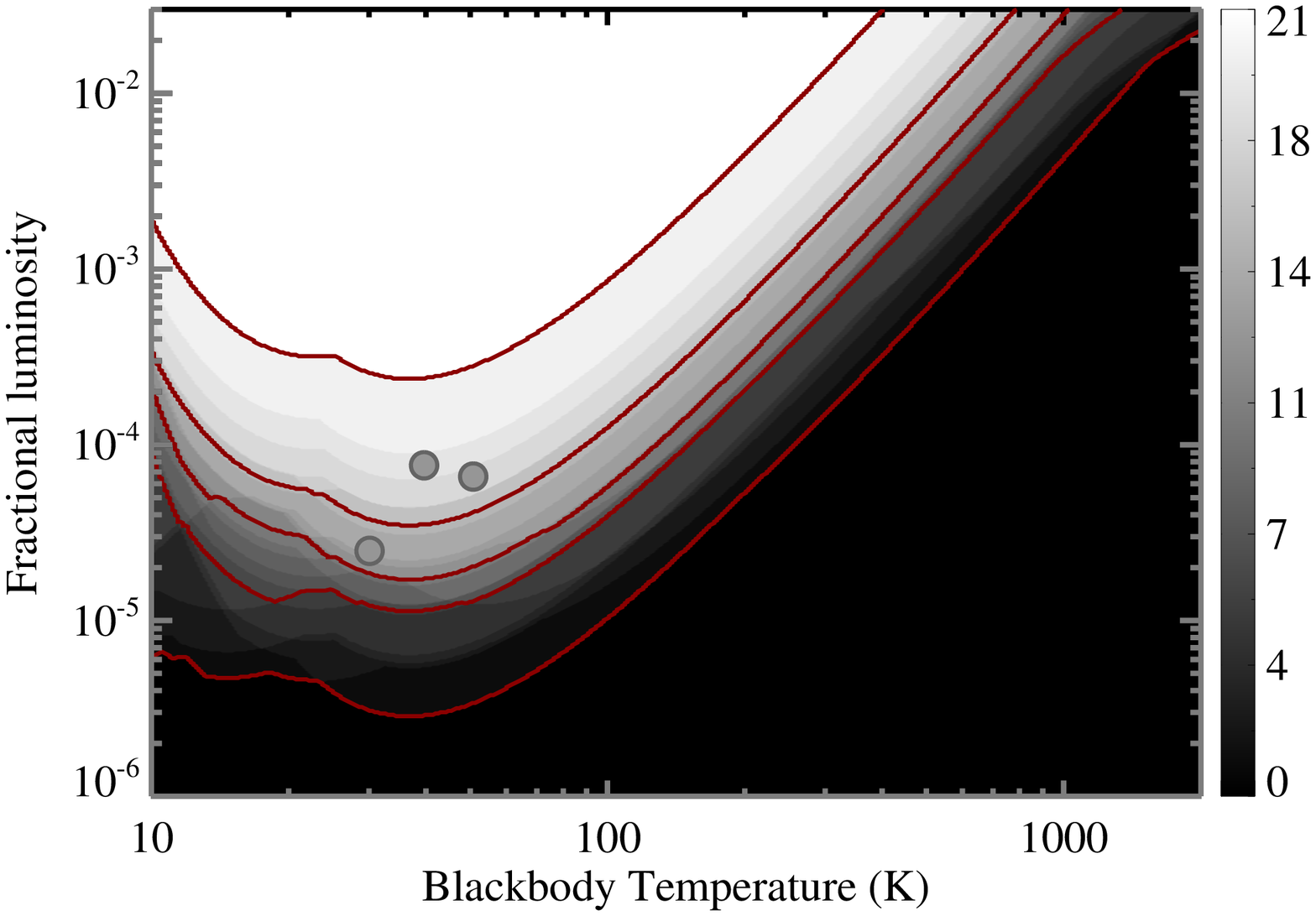}
    \hspace{-0.cm} \includegraphics[width=0.48\textwidth]{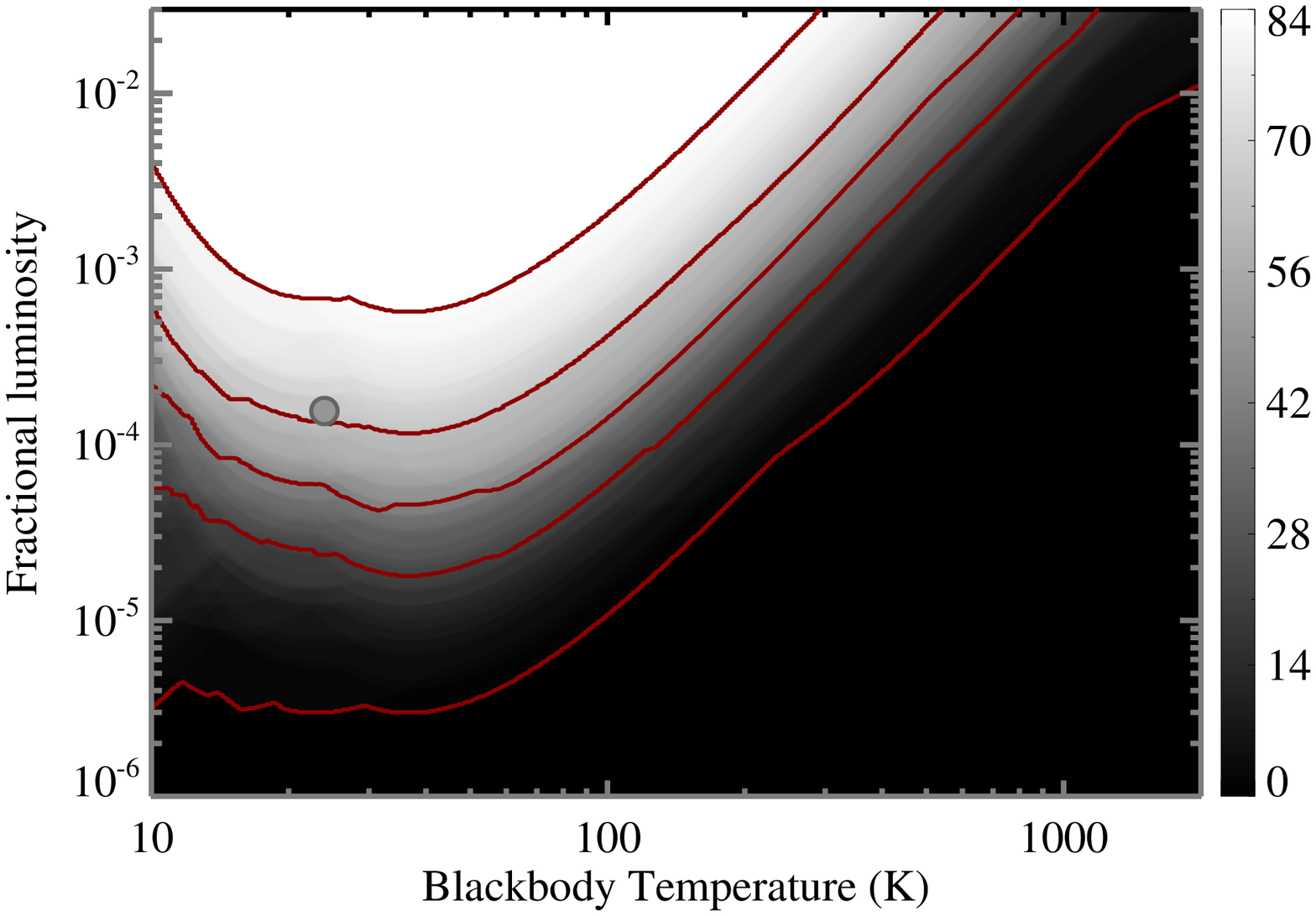}
    \caption{Detection space for our sample (\emph{left panel}) and the
      control sample (\emph{right panel}). Contours show the number of
      stars for which a disk of a given fractional luminosity and
      temperature could have been detected. The upper and lower red
      contours show where disks around all, and one, systems could have
      been detected. The intermediate curves are for 75, 50, and 25\% of
      systems. The difference in sensitivity between our sample and the
      DEBRIS control sample is a factor of a few.}\label{fig:detlims}
  \end{center}
\end{figure*}

Thus, we find suggestive evidence that debris disks are more easily
detected around M-type stars that also host planets. A further
consideration however is whether the observations are biased towards
detections for the planet-host sample. This might be expected given that
our noise level is about half that of the DEBRIS observations of the
control sample, but might also be balanced by the fact that all DEBRIS
M-type stars are within 10pc, and thus on average closer than our
planet-host stars.

The relative sensitivities for the two samples is shown in Figure
\ref{fig:detlims}, where the grey scale shows the number of systems for
which disks at a given temperature and above a certain fractional
luminosity ($f = L_{\rm disk}/L_\star$) could have been detected. The
lowest red contour shows the maximum sensitivity (disks that could have
been detected around only one star), the highest shows the level above
which disks could have been detected around all stars, and the
intermediate contours show where disks could have been detected around
25, 50, and 75\% of systems. By comparing the red contours it can be
seen that our observations could typically detect disks that are a
factor of two to three lower in fractional luminosity than those
observed by DEBRIS (as expected from observations that are 2-3 times
deeper). While the three disks around planet-host stars could have been
detected around 75\% of our sample, they could only have been detected
around about 30\% of the DEBRIS sample. Thus, the evidence for any
correlation between planets and debris disk brightness is weaker than
suggested by the $p$-value above.

The significance of the $p$-value may be further reduced by future
radial velocity observations, because an implicit assumption is that the
stars in the control sample do not host planets in a similar parameter
space range as those around our planet-host sample. This is unlikely to
be true because not all systems in our control sample will have been
observed in search of planets, and our control sample is best termed
`stars with no known planets'. If any of the systems in the control
sample that do not host disks were in fact found to host planets, the
significance of our result would decrease further. If however
Fomalhaut~C were found to host a planet (and a search may be well
motivated by our results), the significance would increase.

As noted earlier, it is not yet known whether M-type stars host a disk
population that is the same or different to those that orbit Sun-like
stars, and a major problem is that obtaining comparably sensitive
observations is challenging. This sensitivity difference can be seen by
comparing the contours in the right panel of Figure \ref{fig:detlims}
with those in Figure 4 of Sibthorpe et al. (2017, MNRAS in press), which
shows the sensitivity for FGK-type stars observed as part of the DEBRIS
survey (and for which an FGK-type disk detection rate of 17\% was
obtained). The 50\% contour for our survey is at best about
$f = 5 \times 10^{-6}$, an order of magnitude better than achieved by
DEBRIS for M-type stars. In comparison, our survey is about midway
between the two in terms of sensitivity. Therefore, with the caveats
that the number of detections is small, and that the results could be
biased by a planet-disk correlation, the fact that we have here obtained
a disk detection rate similar to that seen for Sun-like stars suggests
that in surveys of equal sensitivity in fractional luminosity the disk
detection rate among Sun-like and M-type stars should be approximately
the same.

\subsection{A marginally resolved disk around GJ 649}\label{ss:649}

\begin{figure*}
  \begin{center}
    \hspace{-0.5cm} \includegraphics[width=0.48\textwidth]{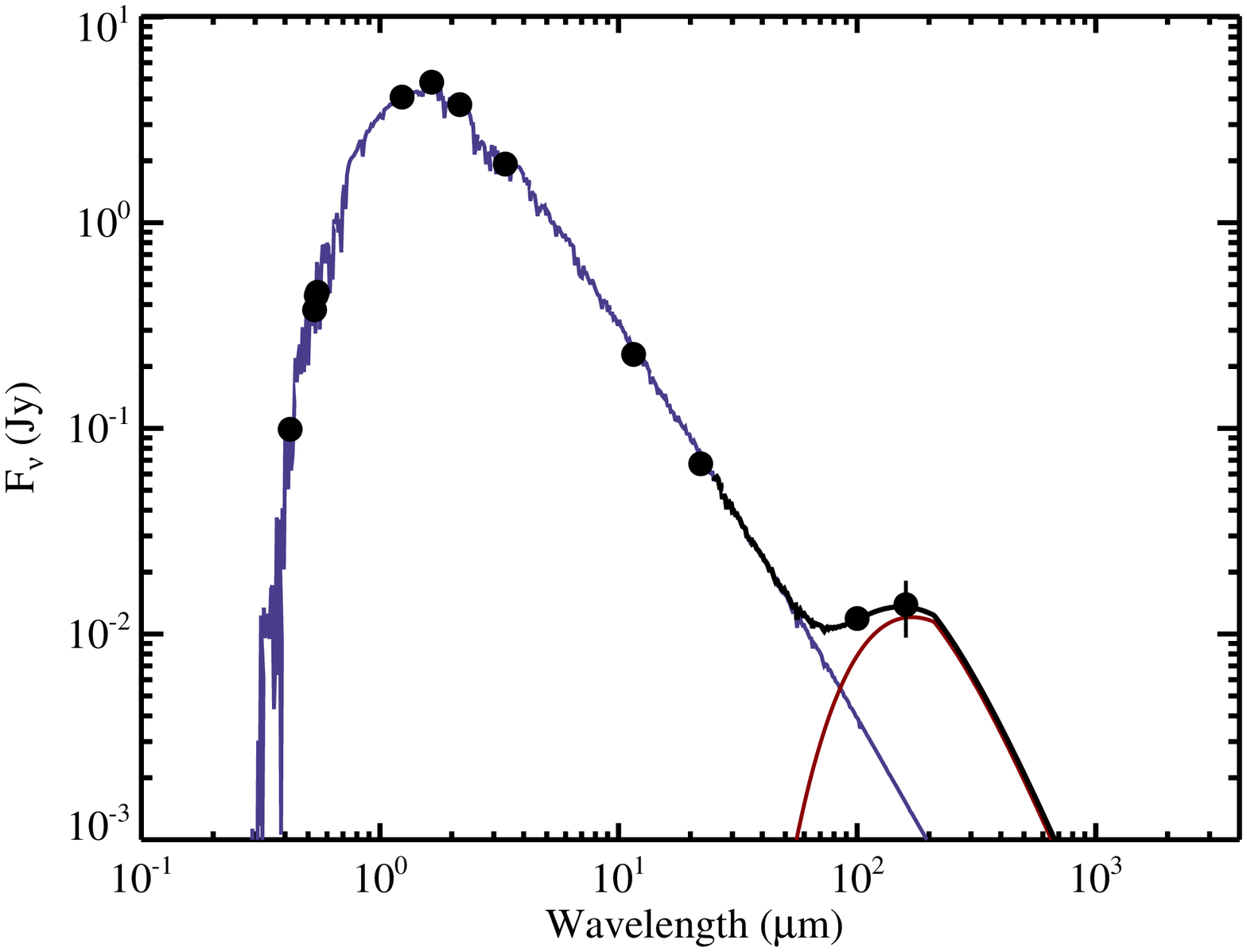}
    \hspace{0.5cm}
    \hspace{-0.5cm} \includegraphics[width=0.48\textwidth]{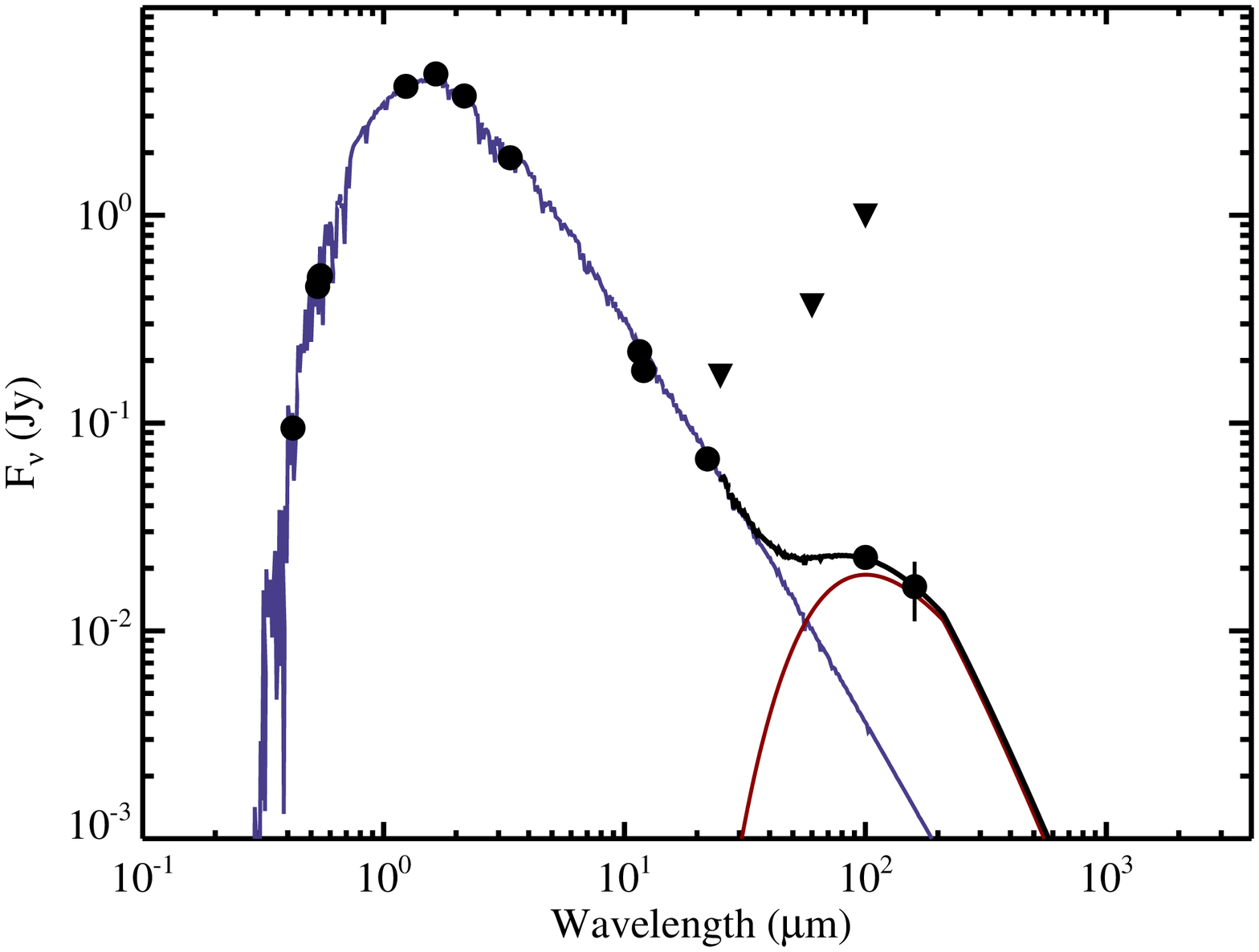}
    \caption{Flux distributions showing the disk detections for GJ~433
      (left panel) and GJ~649 (right panel). Solid lines show the star
      (blue), disk (red), and total (black) models. Black dots and
      triangles show measured photometry and upper limits. The best fit
      disk temperatures are 30 and 50K, though the large uncertainties
      in the 160$\mu$m measurements make these very
      uncertain.}\label{fig:seds}
  \end{center}
\end{figure*}

GJ~649 (HIP 83043, BD+25~3173, LHS 3257) was reported to host a planet
with a minimum mass similar to Saturn's, in an eccentric 598 day (1.1au)
orbit \citep{2010PASP..122..149J}. The age of the star is uncertain,
though it was classed as a member of the `old disk' (as opposed to the
young disk or halo) based on kinematics \citep{1992ApJS...82..351L}, and
noted to be among the 20\% most chromospherically active early M-type
stars \citep{2010PASP..122..149J}. Using constraints from the disk
temperature and \emph{Herschel} images we can therefore build a picture
of the system's architecture.

The flux density distribution for GJ~649 is shown in Figure
\ref{fig:seds}. The excess flux above the photosphere is modelled using
a modified blackbody function, where the disk spectrum is divided by
$\lambda/210\mu{\rm m}$ beyond 210$\mu$m. This steeper long-wavelength
spectral slope approximates the poor efficiency of dust emission at
wavelengths longer than the grain size, though in this case is not
constrained and included simply in order to make the extrapolations to
millimeter wavelengths more realistic. The main point to take away from
this figure is that the dust thermal emission is very cold, so could not
have been detected in the WISE observations at 22$\mu$m. The best-fit
disk temperature is $50$K with $f = 7 \times 10^{-5}$, but is uncertain
because the 160$\mu$m observation is not formally a 3$\sigma$ detection
of the disk (i.e. Table \ref{tab:fluxes} shows that $\chi_{160}$ for
GJ~649 is 2.9). The non-detection of an excess at 22$\mu$m means that
the temperature cannot be significantly more than 100K.

\begin{figure}
  \begin{center}
    \hspace{-0.cm} \includegraphics[width=0.45\textwidth]{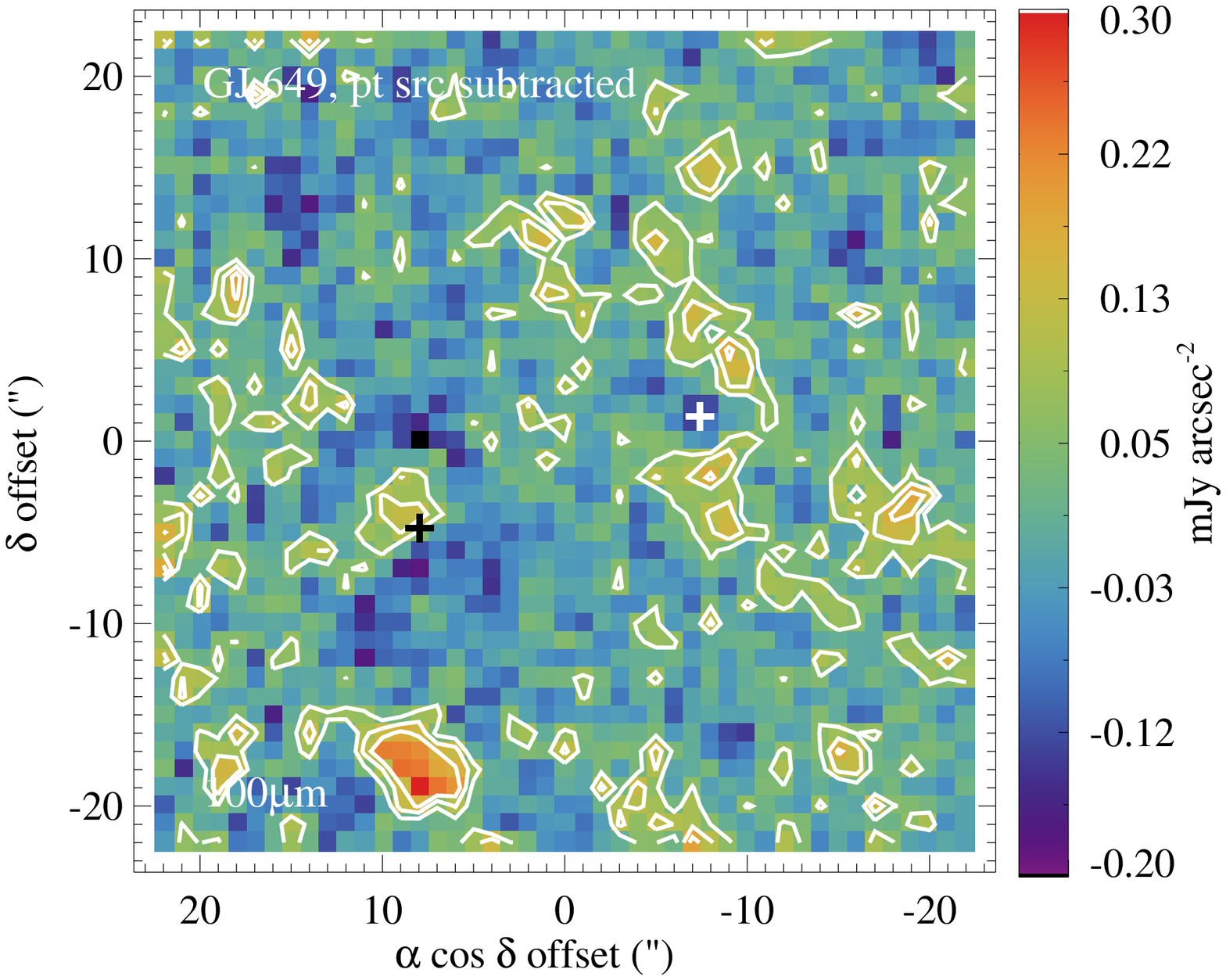}
    \caption{\emph{Herschel} image of GJ~649 after subtracting point
      sources near the location of GJ~649 (at the white +) and at the
      bright peak to the SE (at the black +, see Figure
      \ref{fig:im}). The low level residual structure around GJ~649
      provides circumstantial, though not conclusive, evidence, that the
      disk is resolved. The asymmetry in the residuals suggests that the
      disk position angle is near to North, and that the disk is closer
      to edge-on than face-on. White contours are at 1, 2, and 3 times
      the 1$\sigma$ noise level. The center of the image is
      approximately midway between the plus symbols.}\label{fig:im2}
  \end{center}
\end{figure}

Given a stellar luminosity of 0.044$L_\odot$ the best fit temperature of
50K corresponds to a radial distance of 6au if the disk material behaves
as a blackbody, while a temperature of 100K yields a distance of about
2au. Given that most debris disks are comprised of dust small enough to
have super-blackbody temperatures, the disk around GJ~649 would be
expected to be larger than blackbody estimates, by a factor of several
at least
\citep[e.g.][]{2012ApJ...745..147R,2013MNRAS.428.1263B,2014ApJ...792...65P,2016ApJ...831...97M}.
This factor was found to be 6-20 for GJ~581 \citep{2012A&A...548A..86L},
with the large uncertainty arising because the disk radius depends on
the square of the temperature. At a distance of 10.4pc
\citep{2016A&A...595A...4L} the GJ~649 disk may therefore have an
angular diameter large enough to be resolved. This extent may be
confirmed by the \emph{Herschel} images, which at 100$\mu$m show some
extended residual emission after PSF subtraction (see Figure
\ref{fig:im2}). The fact that these residuals are extended in a
non-axisymmetric pattern suggests that the disk may be nearer to edge-on
than face-on, as might be expected given in the case of a planet
detection with the radial velocity technique. Given that most of the
residual contours are only 1$\sigma$ however, we consider that these
residuals provide circumstantial evidence that the disk is resolved, in
which case the disk diameter would be similar to the PACS beam size of
6\arcsec. We therefore conclude that the disk radius could lie in the
range 2-50au, but is more likely to be a few tens of au.



The system layout is shown in Figure \ref{fig:sys}, where the planet
GJ~649~b is indicated by the dot, and the error bar indicates the range
of radii covered by the eccentric orbit. The solid line shows limits
estimated based on the radial velocity residuals once the best-fit
planet orbit is subtracted,\footnote{The inner part of this limit can be
  derived using Kepler's laws and the residual noise in the RV data once
  the planet(s) have been subtracted, but the steeper outer part where
  the orbital period is longer than the span of observations was
  empirically estimated from full simulations of radial velocity
  sensitivity \citep[e.g.][]{2015MNRAS.449.3121K}} indicating that
planets more massive than Saturn that orbit beyond about 5au would not
have been detected. The range of estimated disk locations is shown by
the hatched region, where we have taken the marginally resolved image to
indicate that the disk has a radius between 10-30au. The basic
conclusion is that while the separation between the planet and disk is
probably large, it is possible that this gap is occupied by one or more
undetected planets. A further conclusion is that lower mass planets at
smaller radii could have been detected, though the sensitivity is a
factor of two poorer than for the other systems discussed below.

\subsection{An unresolved disk around GJ~433}\label{ss:433}

GJ~433 (HIP~56528, LHS~2429) was reported to host a low-mass planet
GJ~433~b ($M \sin i = 5.8 M_\oplus$) on a 7.4 day period at 0.058 au
\citep{2013A&A...553A...8D}. They detected an additional significant
signal with a much longer period of 10 years (3.6au), but based on the
variation of activity indices on a similar timescale
\citep{2011A&A...534A..30G}, concluded that a magnetic cycle of the star
was a more likely origin. The same signals were recovered by
\citet{2014MNRAS.441.1545T}, who considered the second signal to be a
candidate planet.  Given the uncertain nature of the outer planet we do
not include it here. The age of GJ~433 is uncertain, but the dynamical,
x-ray, and Ca~II emission properties show that the star is not young
\citep{2013A&A...553A...8D}.

As above we can constrain the disk location relative to the planet's,
but in the case of GJ~433 there is no clear evidence that the disk is
resolved with \emph{Herschel}. The best fit disk temperature is $30$K
(see Figure \ref{fig:seds}, but again the temperature is poorly
constrained by a weak detection at 160$\mu$m, and could be as warm as
100K. The fractional luminosity is also poorly constrained, but is
approximately $2.5 \times 10^{-5}$. For the stellar luminosity of
0.033$L_\odot$ a disk temperature range from 100 to 30K yields a
blackbody radius range of about 1 to 16au, or about 0.2 to 3.5\arcsec\
diameter at the 9.1pc distance of the system. As for GJ~649, the disk
structure as seen at 100$\mu$m can constrain the disk extent to less
than the PACS beam size, but as with GJ~649 only limits the disk radius
to less than about 30au, and does not constrain the inclination or
position angle.

The system layout is shown in Figure \ref{fig:sys}. While the
observational limits on the disk radius are poor, a radius of 1au would
make GJ~433 host to an unusually small disk \citep{2007ApJ...658..569W},
so it seems most likely that the disk extent is similar to that expected
for GJ~649. If this is indeed the case, there is again space for
undetected planets in the region between the known planet and the disk.

\subsection{Summary of system architectures}\label{ss:arch}

\begin{figure}
  \begin{center}
    \hspace{-0.5cm} \includegraphics[width=0.48\textwidth]{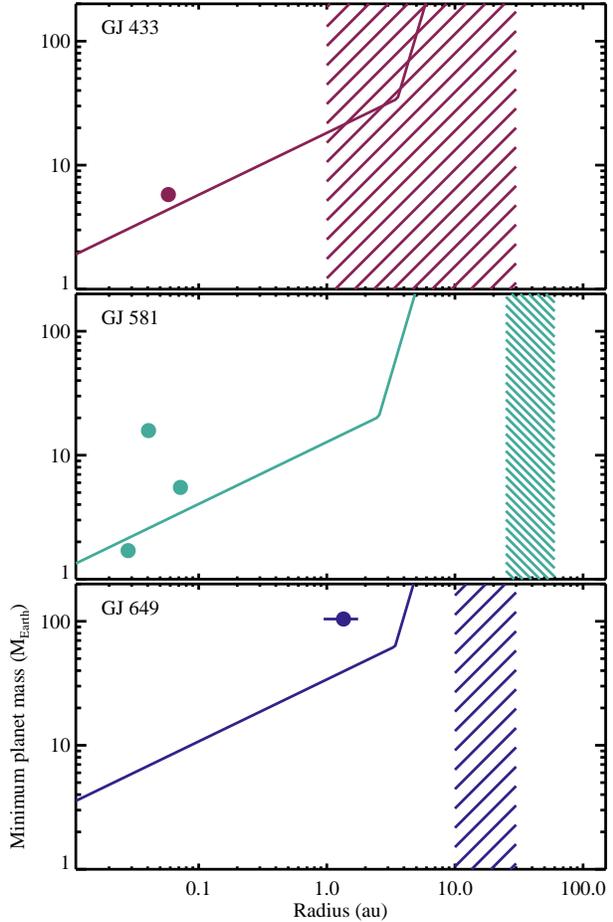}
    \caption{Mass semi-major axis diagrams showing the GJ~433, GJ~581,
      and GJ~649 planets (dots), the approximate RV sensitivity (lines),
      and the possible range of disk locations (hatched regions, showing
      the disk extent in the case of GJ~581). GJ~581~e lies below the
      sensitivity curve because the RV amplitude (1.7 m s$^{-1}$) is
      smaller than the RMS (2.12 m s$^{-1}$) reported by
      \citet{2014Sci...345..440R}. In each case, with the possible
      exception of GJ~433, there remains room in the detection space for
      sizeable planets that reside between the known planets and the
      disk, but that could not have been detected with the current RV
      observations.}\label{fig:sys}
  \end{center}
\end{figure}

Figure \ref{fig:sys} summarises the architecture of the planet-host
systems in our sample, and includes the multi-planet system GJ~581. The
number of planets residing in this system is contentious, and stellar
activity has been proposed as the cause of some of the periodic signals
seen; here we show the three planets proposed by
\citet{2014Sci...345..440R}, and the hatched disk region shows the
extent of the disk derived by \citet{2012A&A...548A..86L}. As with
GJ~433 and GJ~649, there is space for undetected planets in the
intervening region.

Given the lack of strong evidence for any correlation between the
presence of planets and debris disk brightness, we should not
necessarily expect clear trends when looking at plots such as Figure
\ref{fig:sys}. We might however note trends that are glossed over by a
simple disk brightness metric, such as tendencies for systems to show
particular architectures or scales. Again noting that a disk as small as
1au around GJ~433 would be very unusual, the radii of the disks is
consistent with being a few tens of au. However, this size is also
inferred for the disk that orbits Fomalhaut~C
\citep{2014MNRAS.438L..96K}, so there is no evidence that this
preference is related to the presence of planets. Indeed, this radius
range is also preferred for disks around FGK-stars, independent of
whether planets are known (Sibthorpe et al. 2017).

There is no obvious link between the disks and the layout of the planets
that orbit closer in, but in each case there remains room in the
detection space for sizeable planets that reside between the known
planets and the disk, but that could not have been detected with the
current RV observations. In this regard the M-type planet + disk systems
appear to be analogues of Sun-like planet + disk systems such as
HD~20794, HD~38858, and 61 Vir
\citep{2012MNRAS.424.1206W,2015MNRAS.449.3121K}. This similarity may
however simply reflect that detecting long period planets takes time,
and that small debris disks grind down to undetectable levels more
rapidly than large ones, and that these biases are present regardless of
the mass of the host star. That is, there may be differences in the
architectures of planetary systems across different spectral types, but
that this difference is in the type or existence of planets that reside
near 10au. For further discussion of planet formation scenarios, we
refer the reader to \citet{2012MNRAS.424.1206W},
\citet{2015MNRAS.449.3121K}, and \citet{2017MNRAS.469.3518M}.

The very cool disk temperatures shown in Figure \ref{fig:seds} make it
clear that progress in our understanding of these disks, and the links
with the planets, can only be made by far infrared and millimeter-wave
observations. The present observations are hindered by the low spatial
resolution of \emph{Herschel}, which means that we are constrained to
estimating disk locations. With no far infrared missions on the near
horizon, and an expectation of sub-mJy disk flux densities, observations
with the Atacama Large Millimeter Array (ALMA) are the main avenue for
progress. These will be challenging, but necessary to obtain further
discoveries, and in cases such as GJ~433, GJ~581, and GJ~649 could
provide higher resolution images that instead of yielding disk location
estimates, will allow the discussion of disk structure.

\section{Conclusions}\label{s:conc}

This paper presents the results of a \emph{Herschel} survey of 21 nearby
late-type stars that host planets discovered by the radial velocity
technique. These observations were obtained with the aim of discovering
new debris disks in these systems, and in search of any correlation
between planet presence and disk brightness.

We report the discovery of two previously undetected disks, residing at
a few tens of au around the stars GJ~433 and GJ~649. The disk around
GJ~649 appears marginally resolved and more consistent with being viewed
edge-on. Despite uncertainty in their radii these disks orbit well
beyond the known planets, and it is possible that other as-yet
undetected planets reside in the intervening regions. The layout of
these systems therefore appears similar to star + disk systems around
Sun-like stars such as HD~20794, HD~38858, and 61~Vir. Estimating the
ages of M-type stars is challenging, but neither star shows evidence of
youth, so there is no evidence that the ages of these stars are special
compared to the rest of the sample.

Including the previously known disk around GJ~581, our sample comprises
three planet hosts with disks, a detection rate of 14\%. While this rate
is higher than for a control sample of M-type stars without reported
planets observed by the DEBRIS survey (1 out of 84 stars), the
difference is only significant at 98\% confidence. This evidence is
further shown to be optimistic, because the observations of the
planet-host sample were somewhat more sensitive to debris disks than
those in the control sample, and because not all systems in the control
sample have been searched for planets (or reported not to have planets
above some detection threshold).

Though this survey represents an improvement over previous surveys of
M-tye stars, the fractional luminosity sensitivity achieved remains
about a factor of three poorer than similar surveys of Sun-like
stars. Nevertheless, the fact that we find disks around 14\% of M-type
stars, in comparison to 17\% of Sun-like stars, provides circumstantial
evidence that there is no difference in their disk populations.

\section{Acknowledgements}

We thank the referee for a useful report. GMK is supported by the Royal
Society as a Royal Society University Research Fellow. This work was
supported by the European Union through ERC grant number 279973 (GMK \&
MCW).

The Digitized Sky Survey was produced at the Space Telescope Science
Institute under U.S. Government grant NAG W-2166. The images of these
surveys are based on photographic data obtained using the Oschin Schmidt
Telescope on Palomar Mountain and the UK Schmidt Telescope. The plates
were processed into the present compressed digital form with the
permission of these institutions.







\end{document}